\documentclass[aps,prl,reprint,amsmath,amssymb,showpacs,nobibnotes,floatfix]{revtex4-1} 
 
\usepackage{graphicx}
\usepackage{bm}     
\usepackage{hyperref} 
\usepackage{xcolor}
\hypersetup{
    colorlinks,
    linkcolor={blue!80!black},
    citecolor={blue!80!black},
    urlcolor={blue!80!black}
}
\usepackage{dcolumn} 
\usepackage{dsfont}    
\usepackage{color}
\usepackage{textcomp,gensymb} 

\begin{document}

\title{Two-Photon Induced Coherence without Induced Emission}

\author{Dong-Gil Im}
\altaffiliation{Present address: Agency for Defense Development, Daejeon, 34186, Korea}
\email{eastgil7902@gmail.com}
\affiliation{Department of Physics, Pohang University of Science and Technology (POSTECH), Pohang, 37673, Korea}

\author{Seung-Yeun Yoo}
\affiliation{Department of Physics, Pohang University of Science and Technology (POSTECH), Pohang, 37673, Korea}

\author{Chung-Hyun Lee}
\affiliation{Department of Physics, Pohang University of Science and Technology (POSTECH), Pohang, 37673, Korea}

\author{Jongheon Suh}
\affiliation{Department of Physics, Pohang University of Science and Technology (POSTECH), Pohang, 37673, Korea}

\author{Yoon-Ho Kim}
\email{yoonho72@gmail.com}
\affiliation{Department of Physics, Pohang University of Science and Technology (POSTECH), Pohang, 37673, Korea}

\date{\today}

\begin{abstract}
At the heart of recent breakthroughs in quantum imaging and spectroscopy utilizing undetected photons lies the quantum optical effect known as induced coherence without induced emission. This fundamental quantum interference effect has unlocked new possibilities in accessing challenging wavelength regimes for advanced imaging and spectroscopic analysis. Despite these advancements, the full spectrum of quantum metrology's capabilities, particularly the enhanced phase sensitivity offered by quantum optical Fock states or N00N states, has yet to be realized. This is due to the fact that, until now, the exploration of induced coherence has been confined to phenomena involving single photons. In this study, we present the observation of two-photon induced coherence without induced emission. This advancement hinges on a two-photon Fock state that creates quantum coherence between pairs of two-photon spontaneous emission amplitudes. The result is a doubling of the interferometric phase modulation compared to what is observed with single photons. Specifically, we show that a phase change $\phi$ applied to undetected 1016 nm near-infrared photons leads to $2\phi$ modulation in the detection of the 632 nm visible photons, verifying two-photon induced coherence without induced emission. These findings pave the way for innovative high-resolution quantum metrological applications leveraging multi-photon induced coherence without induced emission.
\end{abstract}

\maketitle
\textit{Introduction.}---Optical coherence, induced through light injection, plays a crucial role in numerous applications across photonics, spectroscopy, metrology, and communications. Such induced optical coherence arises from induced emission triggered by the injection of light \cite{Tang67,Stover66,Sergienko}. Surprisingly, induced coherence without induced emission is possible in quantum optics \cite{Zou91,Wang91}, and it arises from the inherent indistinguishability of the probability amplitudes involved in the process \cite{Kwiat94,Kim00,Kim05}. The phenomenon of induced coherence without induced emission is perhaps one of the most fascinating fundamental quantum optical interference effects, as it is uniquely observed in the first-order in intensity \cite{Grayson94_phase,Grayson94_spatial,Kim00b,Lahiri19}, unlike most other quantum optical interference effects, which are exhibited in the second-order intensity-intensity correlation measurement \cite{HOM,shih,Kim99,Kim00a}. In addition to the fundamental importance and interest, this phenomenon forms a basis of nonlinear interferometry in quantum optics \cite{burlakov97,Kory,kulik04,Chekhova16}. 

Recently, the phenomenon of induced coherence without induced emission has been leveraged to make significant breakthroughs in quantum optical metrology utilizing undetected photons, opening up new possibilities in accessing challenging wavelength regimes, e.g., mid-infrared wavelengths, for quantum imaging  \cite{Lemos14,Lahiri15,Kviatkovsky20,Santos22,Vega22,Fuenzalida23}, correlation imaging \cite{Pittman95,Strekalov95,Oh13,Yang23}, quantum spectroscopy \cite{Kalashnikov16,Paterova17,Paterova18,Paterova20}, quantum optical coherence tomography \cite{Paterova18QOCT,Valles18}, and quantum polarimetry \cite{Paterova19}. 
In spite of these advancements, quantum-enhanced phase sensitivity, the hallmark effect in quantum metrology, has not been demonstrated for quantum metrology with undetected photons. This is because the exploration of induced coherence without induced emission has been limited to phenomena involving single photons. 

In this study, we propose and experimentally demonstrate two-photon induced coherence without induced emission. This is achieved by employing a two-photon Fock state in a nonlinear interferometer to induce quantum coherence among pairs of two-photon spontaneous emission amplitudes. The outcome is a doubling of the interferometric phase modulation compared to the modulation observed with single photons in quantum interferometry with undetected photons. Specifically, we demonstrate that applying a phase change $\phi$ to undetected 1016 nm near-infrared photons results in a $2\phi$ modulation when detecting the 632 nm visible photons in our nonlinear interferometer, confirming the presence of two-photon induced coherence without induced emission. 

\begin{figure}[t]
\centering
\includegraphics[width=3.2in]{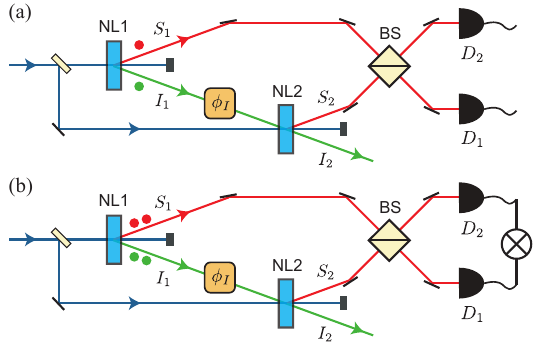}
\caption{(a) Ordinary induced coherence without induced emission is observed at detector $D_1$ or $D_2$ as single-photon interference. 
(b) Proposed two-photon induced coherence without induced emission. The effect is observed in coincidence between detectors $D_1$ and $D_2$.  
}\label{fig1}
\end{figure}

\textit{Conceptual scheme.}---Consider the conceptual schematic of the nonlinear interferometer implementing induced coherence without induced emission in Fig.~\ref{fig1}. The scheme for the ordinary induced coherence without induced emission is shown in Fig.~\ref{fig1}(a) \cite{Zou91,Wang91}. Two identical second-order nonlinear crystals NL1 and NL2 are pumped simultaneously by the same uv pump laser, causing the generation of a pair of signal and idler photons,  $S_1$-$I_1$ from NL1 or $S_2$-$I_2$ from NL2, via spontaneous parametric down-conversion (SPDC). Since the photon-pair of SPDC is in a pure state, the signal ($S_1$ or $S_2$) and the idler photons  ($I_1$ or $I_2$) individually are in thermal states \cite{yurke,strekalov}, and therefore, no first-order interference is expected to occur by overlapping the two signal or the two idler modes. However, when the idler modes $I_1$ from NL1 is overlapped spatio-temporally with $I_2$ through NL2 as shown in Fig.~\ref{fig1}(a), non-classical first-order interference occurs for the signal photon at detector $D_1$ or $D_2$ \cite{Kim00b,Lahiri19}. Surprisingly, the phase modulation $\phi_I$ introduced to the undetected idler photon is exhibited in the interference of the signal photon. If we choose the wavelength of the undetected idler photon in the near-infrared regime where photon detection is challenging, this nonlinear interferometer allows us to use the visible signal photon to measure the phase modulation $\phi_I$. It is important to note that, while this scheme opens new possibilities in accessing challenging wavelength regimes for quantum metrology, quantum-enhanced phase sensitivity cannot be achieved because this is still a single-photon effect.  

Consider now the possibility of multi-photon induced coherence without induced emission, in which $N$ signal-idler photon pairs could be generated either at NL1 or at NL2. The quantum state, in this case, can be written as,
$ 
|\Psi_N\rangle = \frac{1}{\sqrt2} 
\left(e^{ i \, N \phi_I}  |N\rangle_{S_1} |N\rangle_{I_1} + |N\rangle_{S_2} |N\rangle_{I_2}\right),
$
where the subscripts refer to the signal and idler modes as labeled in Fig.~\ref{fig1}(b). When $N$ idler photons are injected into NL2 without induced emission, the idler modes become identical and quantum coherence is induced between the probability amplitudes for the signal photon $|N\rangle_{S_1}$ and $|N\rangle_{S_2}$. Thus, the phase change $\phi_I$ is applied to the undetected idler photon as in Fig.~\ref{fig1}(b), the signal photon should exhibit $N \phi_I$ phase modulation or quantum-enhanced phase sensitivity \cite{Afek10,YSK11,Ra13,Pang14,Hong21,YKim22}, which is the hallmark effect in quantum metrology. Note that the significance of the induced coherence approach to quantum metrology is partly rooted in the fact that the signal photon could be in the visible regime where the detection is simple and efficient, while the idler photon could be in challenging wavelength regimes, such as near-infrared to mid-infrared.


\textit{Experimental results.}---We now describe the experimental demonstration of two-photon induced coherence without induced emission based on SPDC, see Fig.~\ref{fig1}(b). The quantum state of two pairs of photons, in this case, can be written as, 
\begin{equation} \label{eq02}
|\Psi\rangle =  \left( 
\frac{1}{2} e^{i 2 \phi_I}  \hat{a}^{\dagger 2}_{S_1}  \hat{a}^{\dagger 2}_{I_1}
+ \frac{1}{2} \hat{a}^{\dagger 2}_{S_2}  \hat{a}^{\dagger 2}_{I_2}
+ e^{i\phi_I}  \hat{a}_{S_1}^\dagger \hat{a}_{S_2}^\dagger \hat{a}_{I_1}^\dagger \hat{a}_{I_2}^\dagger
\right) |0\rangle,
\end{equation}
where the normalization constant is omitted. For SPDC, when considering the probability amplitudes for two pairs of photons from two sources (NL1 and NL2), it is necessary to consider the probability amplitude for each source contributing a single photon pair, and this is represented in the third term in Eq.~(\ref{eq02}). This term does not carry the desired phase factor of $2 \phi_I$, and we can exclude this term from contaminating our desired $2\phi_I$ modulation in the signal photon via the coincidence measurement between detectors $D_1$ and $D_2$ located at the output modes of BS, as shown in Fig.~\ref{fig1}(b). The identical single photons in modes $S_1$ and $S_2$ are overlapped at BS, and due to the Shih-Alley/Hong-Ou-Mandel interference \cite{HOM,shih}, the third term in Eq.~(\ref{eq02}) only leads to the two-photon coalescence event either at $D_1$ or $D_2$, but never at the coincidence measurement. Thus, assuming perfect matching of the undetected photon modes $I_1$ and $I_2$, the joint detection probability $\mathcal{P}$ between $D_1$ and $D_2$ reveals the two-photon induced coherence without induced emission, exhibiting the quantum-enhanced phase sensitivity of $2\phi_I$, 
\begin{equation} \label{eq03}
\mathcal{P} = 2 + 2\cos (2\phi_I).
\end{equation}

\begin{figure}[t]
\centering
\includegraphics[width=3.2in]{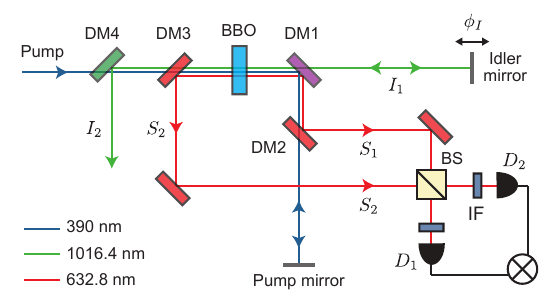}
\caption{Experimental setup corresponding to the schematic shown in Fig.~\ref{fig1}(b). The idler modes $I_1$ and $I_2$ are overlapped by reflecting $I_1$ with the Idler mirror. DM: dichroic mirror, IF: interference filter, BS: beam splitter, 
}\label{fig2}
\end{figure}

\begin{figure}[t]
\centering
\includegraphics[width=3.2in]{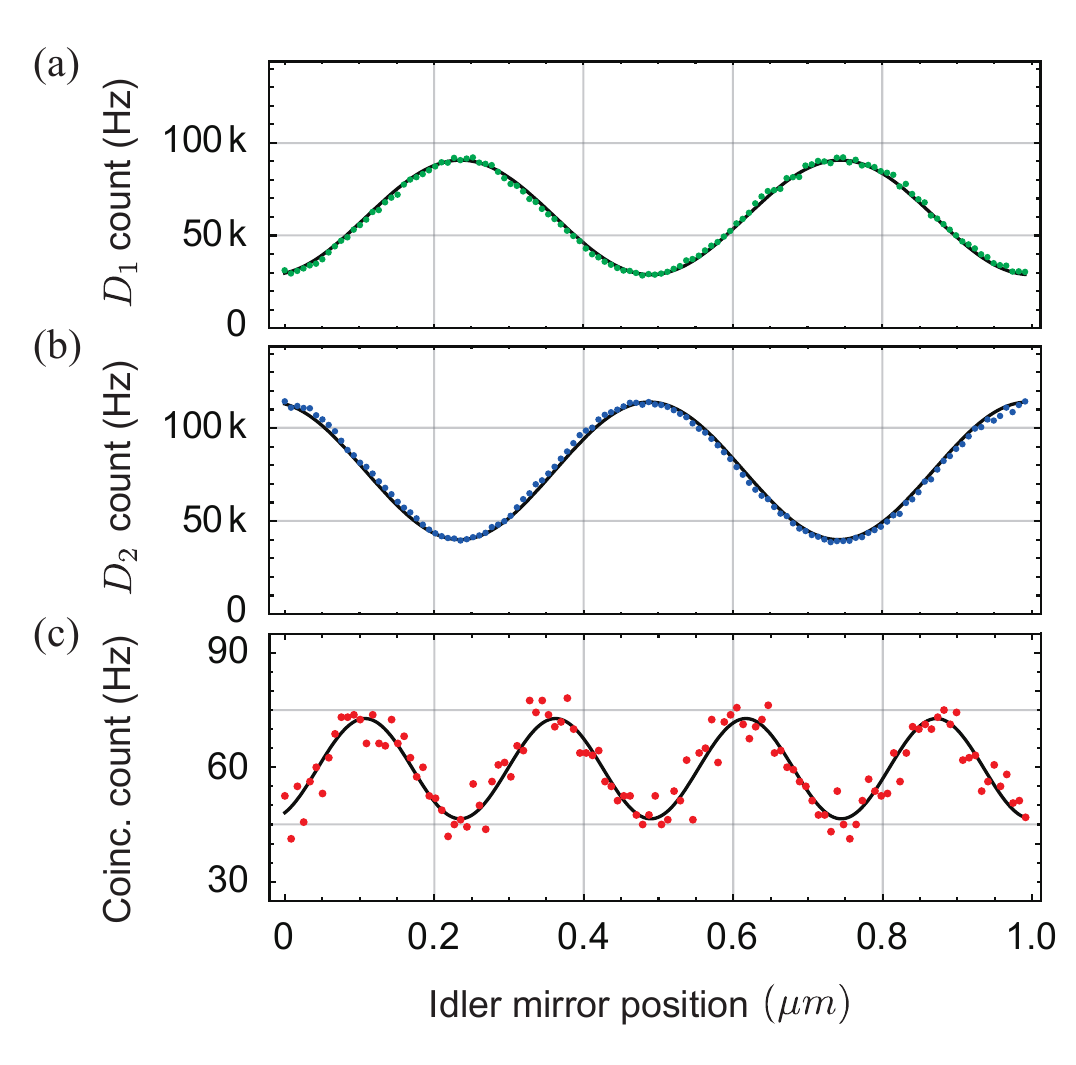}
\caption{Experimental results. Quantum interference at $D_1$, (a), and $D_2$, (b) corresponds to the usual induced coherence without induced emission.  
(c) Two-photon induced coherence without induced emission observed in coincidence counts. The signal photon exhibits $2 \phi_I$ phase modulation, the hallmark effect for quantum-enhanced sensing in quantum metrology. The solid lines represent the sinusoidal fits to the data points.
}\label{fig3}
\end{figure}

The experimental setup is schematically shown in Fig.~\ref{fig2}. An ultrafast femtosecond pulsed laser (Coherent Chameleon) having 390 nm central wavelength is used to pump a 1-mm thick collinear type-II BBO crystal for signal-idler photon pair generation via SPDC. The signal photon wavelength is 632.8 nm, and the undetected idler photon wavelength is 1016.4 nm. We employ a double-pass setup in which the residual forward pump pulse is retro-reflected back to the BBO crystal. The forward and backward SPDC processes in Fig.~\ref{fig2} correspond to NL1 and NL2 in Fig.~\ref{fig1}(b). The idler mode $I_1$ from the forward SPDC, corresponding to NL1 in Fig.~\ref{fig1}(b), is retro-reflected to the BBO to overlap with the $I_2$ mode of the backward SPDC, corresponding to NL2 in Fig.~\ref{fig1}(b), with the Idler mirror. The Idler mirror is controlled with a piezo actuator to add phase $\phi_I$ to the undetected idler photons. The signal modes $S_1$ and $S_2$ from the forward and backward SPDC, respectively, are overlapped at BS and two detectors $D_1$ and $D_2$ are positioned at the two outputs of BS. 
Since we make use of type-II SPDC for the photon pair generation, the signal and idler photons of SPDC have tight frequency anti-correlation \cite{kimgrice}. This results in a mixed state for the quantum state of the signal photon, reducing the visibility of the Shih-Alley/Hong-Ou-Mandel interference in multi-photon quantum interference experiment \cite{mosley,kaneda,wang}. To remove the frequency anti-correlation, the 632.8 nm signal photons are passed through interference filters having the bandwidth of 2-nm at full-width at half-maximum before detection. See Supplementary Materials for details of the joint spectral intensity. Finally, the single-detection and the joint-detection events from the two detectors are recorded as a function of the Idler mirror position.  

\begin{figure}[t]
\centering
\includegraphics[width=3.2in]{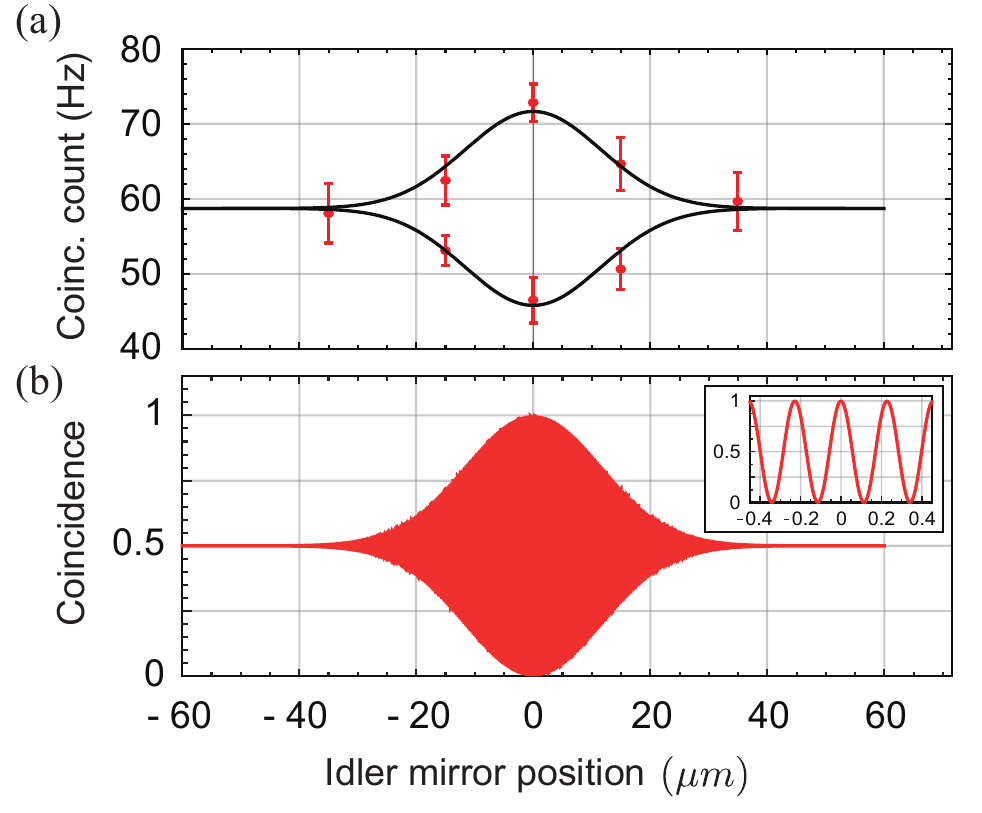}
\caption{Effects of temporal mode mismatch. (a) Maximum and minimum coincidence counts are recorded for a few different Idler mirror positions. The error bars represent one standard deviation. The solid lines represent the Gaussian fits to the data points, showing the envelope of the two-photon induced coherence. 
(b) Numerical simulation of two-photon interference fringes, showing the full envelope of the two-photon induced coherence. The inset shows the interference fringe at the center, exhibiting a period of 508.2 nm.
}\label{fig4}
\end{figure}

The main experimental results are shown in Fig.~\ref{fig3}. As the Idler mirror is translated using a piezo actuator, adding the phase factor $\phi_I$ to the undetected 1016.4 nm idler photons, detectors $D_1$ and $D_2$ registering the 632.8 nm signal photons exhibit the first-order single-photon quantum interference, see Fig.~\ref{fig3}(a) and Fig.~\ref{fig3}(b). These data correspond to the usual induced coherence without induced emission. Note that, since $\phi_I$ is added to the undetected 1016.4 nm idler photon, detectors $D_1$ and $D_2$ registering the 632.8 nm signal photons exhibit the modulation period of 1016.4 nm. The observed quantum interference visibilities are  51.7\% and 47.8\% for Fig.~\ref{fig3}(a) and Fig.~\ref{fig3}(b), respectively. 

The experimental data for two-photon induced coherence without induced emission are shown in Fig.~\ref{fig3}(c). As mentioned earlier, the coincidence event between detectors $D_1$ and $D_2$ behind BS in Fig.~\ref{fig1}(b) picks up only the two-photon induced coherence terms in Eq.~(\ref{eq02}) as the single-photon term is canceled out due to the Shih-Alley/Hong-Ou-Mandel interference at the BS. In this case, the signal photons exhibit $2\phi_I$ phase modulation compared to the case of single-photons, evidenced by the interferometric fringe period of 508.2 nm. The experimental data clearly demonstrate quantum-enhanced sensitivity of two-photon induced coherence without induced emission. 
The two-photon interference visibility is calculated to be 22\%. Somewhat lower two-photon interference visibility is attributed to a slight mode mismatch between the idler modes $I_1$ and $I_2$, stemming from a number of factors, including spatial mode mismatch, differences in beam divergence, channel loss due to non-ideal transmittance/reflectance of the mirrors, etc. 
Here, we emphasize that our two-photon induced coherence without induced emission effect is distinguished from the classical data post processing where multiplying two detector output signals classically. See Supplementary Materials for details.

The effect of temporal mode mismatch is also explored, and the data are shown in Fig.~\ref{fig4}(a) \cite{Kim03,OKwon09,Ihn17,GHL20,CHL23}. The Idler mirror in Fig.~\ref{fig2} is set at five different positions, namely, $-35~ \mu$m, $-15~ \mu$m, $0~ \mu$m, $15~ \mu$m, and $35~ \mu$m, and the maximum and minimum coincidence counts of the two-photon interference near these positions are recorded. The results show the envelope of the two-photon interference with respect to the balanced position of the Idler mirror because the visibility of the two-photon interference decreases as the Idler mirror is positioned away from the balanced position. In other words, two-photon induced coherence is weakened due to the temporal mismatch, which introduces distinguishability. The result of the full multi-mode numerical simulation is plotted in Fig.~\ref{fig4}(b), and is in good agreement with the experimental data in Fig.~\ref{fig4}(a). The inset in Fig.~\ref{fig4}(b) shows the two-photon interference fringe near the center position of the Idler mirror, confirming $2\phi_I$ phase modulation or the period of 508.2 nm. See Supplementary Materials for details of the calculation.

\begin{figure}[t]
\centering
\includegraphics[width=3.4in]{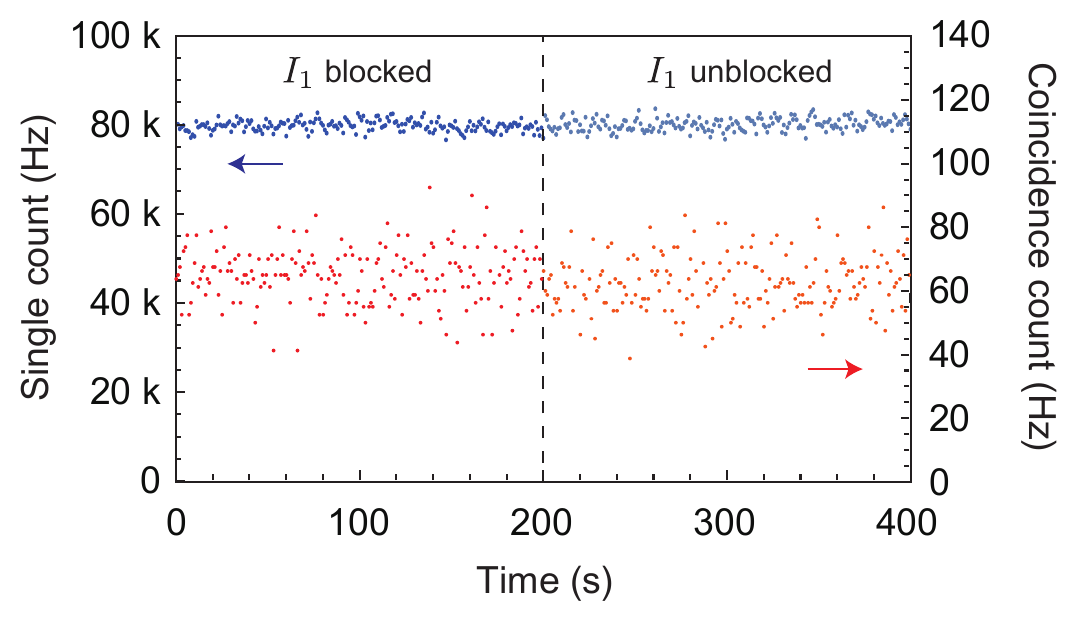}
\caption{Single count and two-photon coincidence count rates in $S_2$ mode before and after unblocking the undetected photon mode $I_1$. The data confirm the absence of induced two-photon emission in our experiment. 
}\label{fig5}
\end{figure}

\textit{Analysis.}---To ensure that induced emission is not present, we place a detector in $S_2$ mode before BS and compare the single-photon count rates before and after unblocking the undetected photon mode $I_1$. If induced emission is present when $I_1$ mode is unblocked, a detector placed in $S_2$ mode will show an increased count rate compared to the case of $I_1$ mode blocked. Initially $I_1$ mode is blocked and the single count rate is continuously measured. At 200 s, $I_1$ mode is unblocked. The experimental data are shown in Fig.~\ref{fig5}. There is no observable increase in the single photon count rate when $I_1$ mode is unblocked, which confirms the absence of induced emission in our experiment. We have also compared the rates of the two-photon event at  $S_2$, before BS with two single-photon detectors and a beam splitter, before and after unblocking the undetected photon mode $I_1$. The coincidence rate also confirms the absence of induced emission in our experiment. See Supplementary Materials for details.

In conclusion, we have proposed, analyzed, and experimentally demonstrated two-photon induced coherence without induced emission. Although induced coherence without induced emission lies at the heart of recent breakthroughs in quantum optical metrology utilizing undetected photons, thereby opening new possibilities in accessing challenging wavelength regimes in various applications, quantum-enhanced phase sensitivity, the hallmark feature of quantum metrology, has not been demonstrated for quantum metrology with undetected photons. This is because the exploration of induced coherence without induced emission has so far been limited to phenomena involving single photons. In our demonstration of the two-photon induced coherence without induced emission, we clearly show a doubling of the interferometric phase modulation compared to what is observed with single photons. Specifically, a phase change $\phi$ applied to undetected 1016 nm near-infrared photons leads to $2\phi$ modulation in the detection of the 632 nm visible photons, verifying quantum-enhanced phase sensitivity corresponding to a two-photon state for quantum metrology with undetected photons.  These findings pave the way for innovative quantum-enhanced metrological applications leveraging multi-photon induced coherence without induced emission.

\section{Acknowledgements}
This work was supported in part by the National Research Foundation of Korea (RS-2023-00208500 and RS-2024-00442762) and the Ministry of Science and ICT of Korea under the Information Technology Research Center support program (IITP-2022-RS-2022-00164799).

\section{Data Availability}
The data presented in this article can be obtained from the corresponding author upon reasonable request.

\section{Conflict of interest}
The authors declare no competing interests.

\section{Supplemental document}
See Supplementary Information for supporting content.





\end{document}